\documentclass[12pt]{article}                            
\usepackage{psfrag}
\usepackage{graphicx}                                                                                

\title{                                                                         
 Monte Carlo simulation of joint density of states in one-dimensional  
Lebwohl-Lasher model using Wang-Landau algorithm
\\[8mm]
} 
\author{Kisor Mukhopadhyay$^a$, Nababrata Ghoshal$^b$ and 
Soumen Kumar Roy$^{c,}$
\footnote{E-mail addresses: $^a$ kisor\_m@yahoo.com, $^b$ ghoshaln@yahoo.co.in,
 $^c$ skroy@phys.jdvu.ac.in,
}\\
$^a$Department of Physics, Sundarban Mahavidyalaya,\\[1mm]
Kakdwip, South 24 Parganas, West Bengal, INDIA\\[1mm]
$^b$Department of Physics, Mahishadal Raj College,\\[1mm] 
Mahishadal, Midnapore (East), West Bengal, INDIA\\[1mm]
$^c$Department of Physics, Jadavpur University,\\[1mm]
Kolkata - 700 032, INDIA} 
\date{  }

\begin{document}

\maketitle
\begin{abstract}
Monte Carlo simulation using the Wang-Landau algorithm has been performed in an 
one-dimensional Lebwohl-Lasher model.  
 Both one-dimensional  and two-dimensional random walks have been carried out. 
The results are compared with the exact results which are available for this 
model.
\end{abstract}
\section{Introduction}
The Wang-Landau (WL) approach \cite{wl} in Monte Carlo (MC) simulation, 
introduced in 2001, has since been applied to different areas of statistical 
physics. While proposing the algorithm the authors have demonstrated the 
application of the algorithm to systems with discrete energy levels (like the 
Ising system, Potts model, spin-glass etc.). Over the past few years several 
authors \cite{lj1,lj2,jaya} have reported the application of the WL method to 
systems with 
continuous energy spectrum. The common feature of these simulations is the 
expected discretization of the energy range which has been investigated {\it 
i.e.} the division of the energy range into a number of bins and the 
subsequent application of the ideas of Wang and Landau. It may however be 
noted that in majority of these simulations a random walk in energy space 
alone was conducted. Relatively few papers \cite{2dwalk,cont} have so far 
appeared where a two dimensional random walk has been performed in a 
continuous system.

It has generally been noted that even for a modest system size, the simulation 
of joint density of states, where the variables for instance are energy and 
the order parameter, (both quantities being continuously variable), the computational time necessary for the implementation of the WL algorithm is very large. 
It may further be noted that till now no work has been published regarding the 
errors involved in the WL-simulation in such systems. Also, no simulation 
has so far been reported where comparison has been made with the exact 
results available (as is usually done for discrete systems with the two 
dimensional Ising model). The reason for this  may 
of course be attributed to the non-availability of results of exact 
calculation for any non-trivial system having a continuous energy spectrum.

The present work has been intended to fill the gap in the availability of 
adequate information in this direction to some extent. The system we have 
chosen for this purpose is an one-dimensional array of three-dimensional spins 
($d=$1, $n=$3, where $d$ is the system dimensionality and $n$ is the spin dimensionality) interacting with nearest neighbours via a potential 
$-P_2 (\cos \theta_{ij})$, where $P_2$ is the second Legendre 
polynomial and $\theta_{ij}$  is the angle between the nearest neighbour spins $i$ and $j$, (the coupling constant in the interaction has been set to unity).
This model, known as the Lebwohl-Lasher (LL) model \cite{ll}, is the lattice 
version of the Maier-Saupe (MS) model \cite{ms1,ms2,ms3} which describes a 
nematic liquid crystal in the mean field approximation. 

The one-dimensional LL model ($d=$1, $n=$3), which of course does not have any ordered state and consequently can not exhibit any finite temperature 
order-disorder transition, has been solved exactly by Vuillermot and Romerio 
\cite{exact1,exact2} 
in 1973, using a group theoretical method. We decided to choose this simple 
model to apply and test the performance of the WL algorithm for simulation of 
joint 
density of states so that a comparison can be made with the exact results  
available. 

We have performed WL simulation using one-dimensional random walk, where the 
visits are confined to the energy space alone, and also a two-dimensional 
random walk where the two dimensional space spanned is the energy-order 
parameter space as well as energy-correlation function space. The partition 
function $Z$ can directly be computed as a function of temperature from a 
knowledge of the density of states, $g(E) = \ln \Omega(E)$, using the relation 
\begin{equation}
Z(T) = \sum_E \Omega(E) e^{-\beta E}
\end{equation}    
where $\beta = 1/T$ (the Boltzmann constant has been set to unity). In the two 
dimensional walk one computes $\Omega (E,\phi)$ where $\phi$ is the order 
parameter of the system or a correlation function which are defined in a 
following section. The partition function can be computed from a knowledge of 
$\Omega(E,\phi)$:
\begin{equation}
Z(T) = \sum_E \sum_\phi \Omega(E,\phi) e^{-\beta E}
\end{equation}
The ensemble average of any function of $\phi$ at a temperature $T$ is given 
by,
\begin{equation}
\langle f(\phi,T) \rangle = \frac{\sum_E \sum_\phi f(\phi) \Omega(E,\phi)  
e^{-\beta E}}{\sum_E \sum_\phi \Omega(E,\phi) e^{-\beta E}}
\end{equation}

We have computed $\ln Z$ from both 1-d and 2-d random walks and have compared 
the results obtained with those available from the exact results of Ref. 
\cite{exact1,exact2}.  
\section{The one dimensional Lebwohl-Lasher model and the exact results}
The Hamiltonian of the Lebwohl-Lasher model is given by
\begin{equation}
H = -\sum_{\langle i,j \rangle} P_2(\cos \theta_{ij})
\end{equation}
where $P_2$ is the second order Legendre polynomial and $\theta_{ij}$ is the 
angle between the nearest neighbour spins $i$ and $j$. The spins are three dimensional 
and headless, {\it i.e.} the system has the O(3) as well as the local $Z_2$ 
symmetry characteristic of a nematic liquid crystal. A vector order parameter 
is inadequate for the system and a traceless second rank tensor {\underline Q},
 as defined below, is used to describe the orientational order of the system 
\cite{lub}. 
One uses, 
\begin{equation}
Q_{ij} = \frac{1}{N} \sum_{t=1}^{N} \left ( n_{i}^{t} n_{j}^{t} - \frac{1}{3} 
\delta_{ij} \right)
\end{equation}
where $n_{i}^{t}$ is the $i$-th component of the unit vector $\hat n$, which 
points along the spin at the site $t$. $N$ is the number of particles in the 
system. In the ordered state $\langle \underline Q \rangle$ is non-zero. 
In a coordinate system with the Z-axis points along the direction of molecular 
alignment (director) the matrix $\langle \underline Q \rangle$ is diagonal and 
for a uniaxial system, 
\begin{equation}
\mathbf{\langle \underline Q \rangle} = 
S \left( \begin{array}{ccc}
-1/3 & 0 & 0 \\
0 & -1/3 & 0 \\
0 & 0 & 2/3 
\end{array} \right )
\end{equation}
where, 

\begin{equation}
S=\frac{1}{2} \langle (3 \cos^2 \theta^t - 1)\rangle = \langle 
P_2(\cos\theta^t)\rangle 
\end{equation}
where $\theta^t$ is the angle between a spin and the director. 

MC simulations demonstrate that a three dimensional Lebwohl-Lasher model 
($d$=3, $n$=3) exhibits a weakly first order transition, 
characteristic of a nematic-isotropic transition which is available from 
the Maier-Saupe model of a nematic in the mean field approximation. On the 
other hand, for lattice dimensionality $d$=2 and 1, no true long range 
order is expected since Mermin-Wagner theorem \cite{mermin} predicts a 
fluctuation destruction 
of long range order. The $d$=2 LL model has been investigated by a number of 
authors \cite{kunz,mondal} and the system shows a behaviour qualitatively similar to the two 
dimensional XY model. A quasi-long range order has been observed in this 
system and this is believed to be related to the existence of topological 
defects in the system \cite{dutta}. 

The one-dimensional Lebwohl-Lasher model has been simulated by \cite{zan} and 
has also been solved exactly \cite{exact1,exact2}. The system is known to be disordered at all 
finite temperatures and critical behaviour is expected only at $T$=0, which 
resembles an one-dimensional Ising model or the one dimensional Heisenberg 
model. The second rank spin-spin correlation 
function $\rho(r)$ is defined as
\begin{equation}
\rho(r) =  \langle P_2 \left ( \cos \theta (r) \right ) \rangle 
\end{equation}
where $\theta(r)$ is the angle between two spins, $r$ lattice spacings apart. 
In the thermodynamic limit one would expect both $S$ and $Lim \; r \to \infty 
\; \rho(r)$ to vanish whereas in finite systems because 
of finite size effects both quantities may appear to have small but finite 
values.

Vuillermot and Romerio \cite{exact1,exact2} presented an exact solution of the planar 
($n$=2) 
and spatial ($n$=3) versions of the Lebwohl-Lasher model in one dimension 
($d$=1) for a nematic liquid crystal, without periodic boundary conditions. 
They also calculated the two-molecule correlation functions and have shown that
 these models do not exhibit any finite temperature  order-disorder phase 
transition. 

The partition function $Z_N(\tilde 
K)$ for the $N$-particle system is given by 
\begin{equation}
Z_N(\tilde K) = \tilde K^{-N/2} \exp[\frac{2}{3}N \tilde K] D^N (\tilde K^{1/2}) 
\end{equation}
where $\tilde K$=3/2T. $D$ is Dawson function \cite{daw}.
$$
D(x) = \exp[-x^2] \int_{0}^{x} du \exp[u^2]
$$
The dimensionless internal energy $u_N (\tilde K)$, the entropy $S_N (\tilde 
K)$ and the specific heat $C_N (\tilde K)$ are given by 

\begin{equation}
\frac{2 U_N (\tilde K)}{N} = 1 + \frac{3 \tilde K^{-1}}{2} - \frac{3}{2} 
\tilde K^{-1/2} D^{-1} (\tilde K^{1/2})
\end{equation}
\begin{equation}
\frac{S_{N} (\tilde K)}{N} = \frac{1}{2} + \tilde K - \frac{1}{2} \tilde 
K^{1/2} D^{-1} (\tilde K^{1/2}) + \ln \left [ \tilde K^{-1/2} D(\tilde K^{1/2}) \right  ]
\end{equation}
and 
\begin{equation}
\frac{2 C_N (\tilde K)}{N} = 1 - \tilde K^{3/2} \left [ \frac{\tilde K^{-1}}{2} 
- 1 \right ] D^{-1} (\tilde K^{1/2}) - \frac{1}{2} \tilde K D^{-2} (\tilde K^{
1/2}).
\end{equation}
The correlation function is given by
\begin{equation}
\rho_N(r) = \left [ \frac{3}{4}  \tilde K^{-1/2}  D^{-1} (\tilde K^{1/2}) - 
\frac{3}{4} \tilde K^{-1} - \frac{1}{2} \right ]^r
\end{equation}

\section{Computational details} 
In the model we have investigated, 
 spins can take up any orientation in the three dimensional space and 
the orientation of each spin is stored in terms of the direction cosines 
($l_1$, $l_2$, $l_3$). The starting configuration has always been chosen as a 
random one and to generate a new microstate, a randomly selected spin is chosen 
and each direction cosine of it is updated as $l_i \rightarrow l_i + p*r_i$ 
(for 
i=1,2,3) where $p$ is a parameter to be chosen according to some criterion and 
$r_i$ is a random number between -1 to +1. To preserve the unit magnitude of 
the spins, ($l_1$, $l_2$, $l_3$) is always normalized. 

The energy of the system in the LL model is a continuous variable and in one dimension ($d=$1) it can have any value between $-L$ to 
$L$/2. To have a discretization scheme for the implementation of the WL 
algorithm and for an one dimensional random walk in the energy space, we have 
chosen an energy range from (-$L$ to 0) and divided this energy range into $M$ 
bins each having a width $d_e$.

We use $g(E_i) = \ln \Omega(E_i)$ where, $\Omega (E_i)$ is the number of 
micro-states corresponding to the $i$-th bin for which the mid-point has the 
value $E_i$. Initially we set all $g(E_i)$ (i=1,M) to zero and the logarithm 
of the modification factor 
$\ln f$ is taken as 1. Whenever a new microstate is generated by rotating a 
spin, 
the new system-energy and hence, the macrostate j is determined. Whether the 
move is accepted or not is decided according to the WL prescription \cite{wl} 
for the probability 

\begin{equation}
p_{i \rightarrow j} = min \left ( \frac{\Omega(E_i)}{\Omega(E_j)}, 1 \right ). 
\end{equation}
If the state j is accepted, we make $g(E_j) = g(E_j) + \ln f$ and $h(E_j) = 
h(E_j)$ +1, where $h(E_j)$ is the histogram count. Otherwise we make $g(E_i) 
= g(E_i) + \ln f$ and $h(E_i) = h(E_i)$ + 1. This procedure is repeated for 
10$^4$ 
MC sweeps (where one MC sweeps consists of $L$ attempted moves) and the 
flatness of the histogram is checked and the cycle is repeated till 
90$\%$ flatness in the histogram is reached. This completes one iteration, 
following which we 
reduce the logarithm of the modification factor $\ln f \rightarrow \ln f/2$, 
reset the histogram, and 
the whole procedure is repeated. For each lattice size we have continued with 
the iterations till $\ln f$ gets reduced to 10$^{-9}$.

We have also calculated the quantity $S$ (Eq. (7)) which gives us the 
magnitude of the order parameter obtained from 
the largest eigenvalue of the ordering matrix defined in Eq. (5) and a 
two-dimensional random walk was performed in the ($E$-$S$) space for this 
purpose. This is necessary 
if one intends 
to determine quantities other than those like free energy, entropy, specific 
heat etc. which are directly related to energy, and is particularly useful, 
if one needs to compute, for instance, the variation of the order parameter in 
presence of an external field.  
It is also possible to calculate the order parameter from an one-dimensional 
walk 
in energy space alone, as has been demonstrated in ref. \cite{jaya} by Jayasri 
{\it et. al.} for a liquid crystalline system (Here one uses the procedure of 
the so called histogram `unweighting' and `reweighting'). But for a more 
accurate calculation of the order parameter (or the correlation function), it 
is perhaps a good idea, to generate a two dimensional walk in the ($E$-$S$) 
space (or in the ($E$-$\rho(r)$) space) and check its flatness.   
 For a given value of the energy of the system, the order parameter, $S$ has a distribution over a certain range of values. The 
whole range of $S$ is 0 to 1  and 
in order to perform the two dimensional random walk in the energy-order 
parameter ($E$-$S$) space we divide the two dimensional space into $M \times 
N$ bins. We represent by $d_\phi$ the bin-width of the bins involving the 
parameter other than energy in the two-dimensional walk. Each microstate will 
now correspond to a macrostate labelled by 
the indices $i$ and $j$ and the acceptance probability given by Eq. (14) is now 
modified to 
\begin{equation}
p_{{ij} \rightarrow {kl}} = min \left ( \Omega(E_i,S_j)/\Omega(E_k,S_l),1\right )
\end{equation}
along with an appropriate modification of the procedure described after 
Eq. (14) for the two-dimensional random walk. Here, for instance, 
$\Omega(E_i,S_j)$ is the density of states for the $i$-th energy and $j$-th 
order parameter bin. 
      
A two-dimensional random walk in the ($E$-$S$) space is a lot more expensive
in terms of the CPU time than an one-dimensional walk in the energy space 
alone. The problem is particularly severe in a system with continuous energy 
and becomes worse as the lattice size increases. However, this ensures a much 
more uniform sampling of the order parameter bins that correspond to a 
particular energy bin and this improves the overall statistics of the work. 
It may be pointed out that 
it is impossible to arrive at a flat histogram in the ($E$-$S$) space if one 
attempts to visit the entire energy and order parameter ranges accessible to 
the system. For an one-dimensional walk one normally faces a problem in that, 
it takes a relatively long time to visit the lowest energy levels and this 
increases with the increase in system size. For a two-dimensional walk the 
possibility 
of uniformly visiting the entire rectangular ($E$-$S$) space is unphysical and 
one must have a prior knowledge of the range of $S$-bins which are likely 
to be visited while the system energy has a given value $E_i$. Our method of 
simulating the two dimensional random walk has resemblance to the work of 
Troster and Dellago \cite{dellago}, who have applied the WL 
algorithm to evaluate multidimensional integrals of sharply peaked functions. 
Our modified approach is elaborated in the following paragraph. 
   
We have first mapped the ($E$-$S$) space which costed us 35 $\times$ 10$^6$
sweeps (to be called the pre-production run). The idea is to determine the 
minimum ($S_{min}^{i}$) and maximum ($S_{max}^{i}$) values of the $S$-bins 
which are visited 
while the system energy is $E_i$ for i=1,M. We observe that there are always 
some $S$-bins within the range ($S_{min}^{i}$,$S_{max}^{i}$) for each $E_i$, 
where either no sampling or very low sampling takes place during the 
pre-production 
run. We therefore checked the histograms of the $(E_i,S_j$) bins in the mapped 
region of the two-dimensional space and those which attain a 90$\%$ flatness 
during the pre-production run are marked with `1' while other bins are 
marked `0'. This may be clarified as follows. We calculate the average 
histogram value for those bins which have been visited at least once, thus 
discarding the bins which are not visited at all. The flatness test (which 
needs each of the visited bins to have a histogram count at least equal to 
90$\%$ of the average histogram) is then applied only to those bins and these 
are labelled with `1'.  In the `production run' 
part of the rest of the simulation we check the flatness of only those bins 
which were marked `1' ignoring what is happening to the others. There is 
however, always a possibility, since the `production run' generates many more 
microstates than that in the `pre-production run', that larger areas in the 
($E$-$S$) space would get included in the initial `visit-map' or those 
bins, once marked `0', would subsequently qualify for the label `1'. But it is 
impossible to improve upon the accuracy of the work indefinitely and we decided 
to stick to the map we obtained during a reasonable amount of the 
`pre-production run', ignoring what is happening to the discarded bins.   
\begin{figure}[tbh]
\begin{center}
\psfrag{ln \Omega(E)}{$\ln \Omega(E)$}
\psfrag{x}{$\Downarrow$}
\rotatebox{-90}{\includegraphics[scale=0.6]{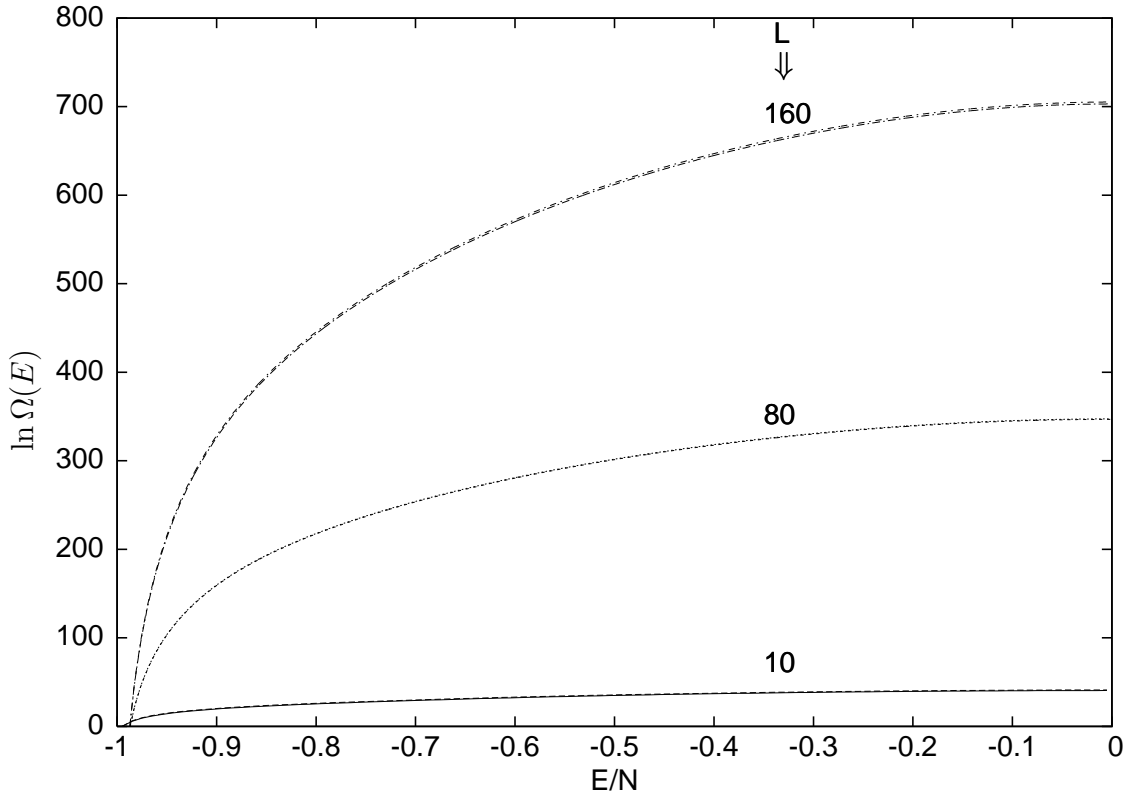}}
\end{center}
\caption{Logarithm of the density of states, $\ln \Omega(E)$, for the 1-d
Lebwohl-Lasher model for $L$=10, 80 and 160 obtained from 1-d and 2-d walks.
In the resolution of the figure the data for 1-d and 2-d walks overlap.} 
\label{1dnnge}
\end{figure}

In addition to the two-dimensional random walk in the ($E$-$S$) space we have also 
performed a number of other two-dimensional random walks. These involve the 
($E$-$\rho(r)$) space where $\rho(r)$ is the correlation function defined in 
Eq. (8). We have done these only for the $L$=160 lattice, for $r$ ranging from 2 to 40 and the ensemble averages of $\rho(r)$ were evaluated for different 
temperatures using Eq. (3).
\begin{figure}[tbh]
\begin{center}
\psfrag{
\psfrag{T}{$T$}
\rotatebox{-90}{\includegraphics[scale=0.6]{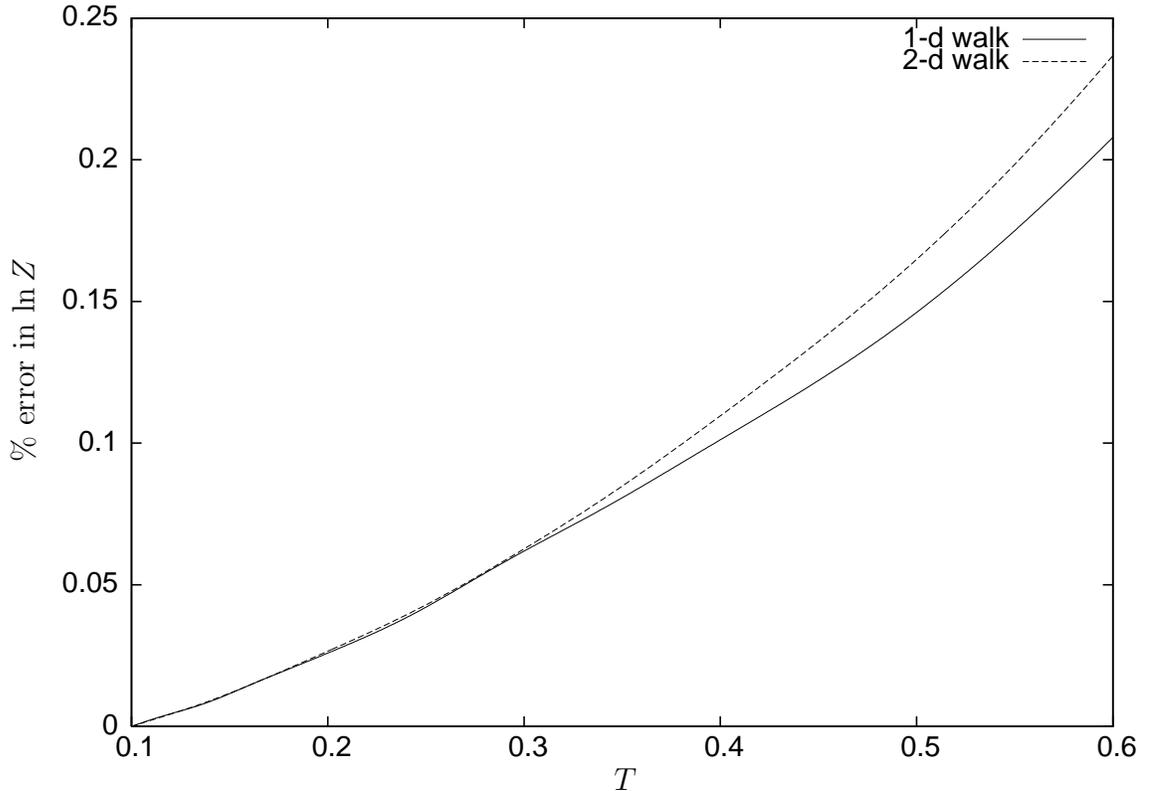}}
\end{center}
\caption{ Percentage error in the logarithm of partition function, $\ln Z$,
for 1-d and 2-d walks,  plotted
against temperature $T$, for the 1-d Lebwohl-Lasher model for $L$=80.}
\label{error}
\end{figure}
\section{Results and Discussions}
We have performed MC simulations using the WL algorithm in linear spin chains 
of length $L$ where $L$=10, 20, 40, 80 and 160. All the results we present 
are results of a single simulation for each lattice size and we did not perform 
averaging of results over multiple simulations although this surely is expected to reduce the errors. In a simulation involving a 
continuous model one is confronted with the proper choice of the values of 
two parameters, $p$ and $d_e$ (defined in Sec. 3). The former determines 
the amplitude of the random rotation of a spin and the latter is the energy 
bin width. In the case of a two dimensional random walk, another parameter
$d_\phi$, which represents the width of the order parameter or the correlation 
function bin is also to be chosen. We have set $d_e$=0.1 for all the work 
reported in this paper. $d_\phi$ was taken to be 0.01 for both order parameter 
and correlation function. The parameter $p$ was always 0.1, except for the two- 
dimensional walk involving the order parameter, where it was taken to be 0.2. 
For larger values of $p$, the CPU time is less, but the results of the 
simulation (like the position and height of the specific heat curve, for 
instance) tend to depend strongly on $p$. For the values of $p$ in the 
neighbourhood of 0.1, the results depend very weakly on $p$. This is 
presumably due to the fact that, a small change in the orientation of a spin 
(one at a time) results in a systematic and uniform sampling in the 
phase space, but as a result of greater correlation of the successive 
configurations generated and the consequent slow movement of the representative 
point in phase space, the computer time involved is greater. For the 
two-dimensional walk involving the order parameter, since a lot of CPU time 
is necessary, we have chosen $p$=0.2, to reduce the time involved. The bin 
widths for energy or other variables were so chosen that for about 50$\%$ of 
the configurations generated by the spin rotation procedure, new bins are 
visited. 
This procedure was found to be optimum, as an attempt to visit new bins more 
frequently, would result in missing a vast majority of the microstates which 
correspond to each bin. A small value of $p$, the rotation amplitude, is 
justifiable from the same point of view. A relatively large value of $p$ 
results in a poor sampling of the infinite number of closely spaced 
microstates contained in each bin and leads to poor results which tend 
to depend on the value of $p$, and consequently not in agreement with the exact 
results.
\begin{figure}[tbh!]
\begin{center}
\psfrag{Cv/N}{$C_v/N$}
\psfrag{T}{$T$}
\rotatebox{-90}{\includegraphics[scale=0.6]{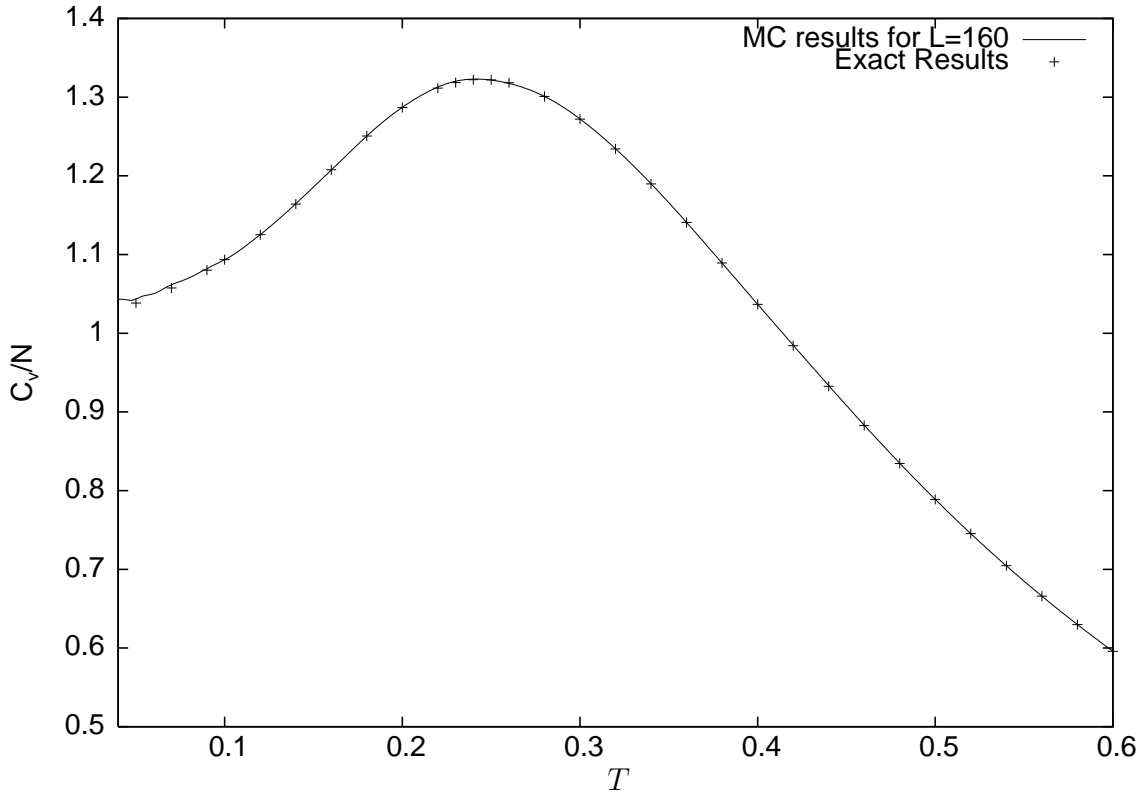}}
\end{center}
\caption{The specific heat per particle, obtained as a fluctuation quantity,
plotted against temperature, for $L$=160 and compared with the exact result.
} 
\label{1dcv}
\end{figure}

In Fig. 1 we have plotted the quantity $g(E)=\ln \Omega(E)$ as a function of 
the energy per particle for $L$=10, 80 and 160 and the results obtained from 
one 
and two dimensional walks (in $E$-$S$ space) have been compared. The system energy was always 
considered upto $E$=0. The lower limit of the energy for $L$=80 was -79 and 
for $L$=160, it was -158,  where the corresponding ground state energies are 
-80 and -160. Thus, the visited energy range goes to a sufficiently low 
value to cover the entire range of interest though the small cut near 
the ground state was necessary, as it takes a huge time to sample these 
states for a relatively 
large lattice. The partition function, $Z$ was calculated from a knowledge of 
the density of states, and the percentage error in $\ln Z$, in comparison 
with the exact results, has been shown in Fig. 2 for both 1-d and 2-d walk 
for the $L$=80 lattice. The error in  $\ln Z$ slowly increases with 
temperature and, at the highest temperature we have investigated, (T=0.6), 
it is about 0.2$\%$; the error in  $\ln Z$ available from 2-d walk being 
marginally higher. 
\begin{figure}[tbh!]
\begin{center}
\psfrag{rho(r)}{$\rho(r)$}
\psfrag{ln[g(E,rho(r))]}{$g(E,\rho(r))$}
\rotatebox{-90}{\includegraphics[scale=0.6]{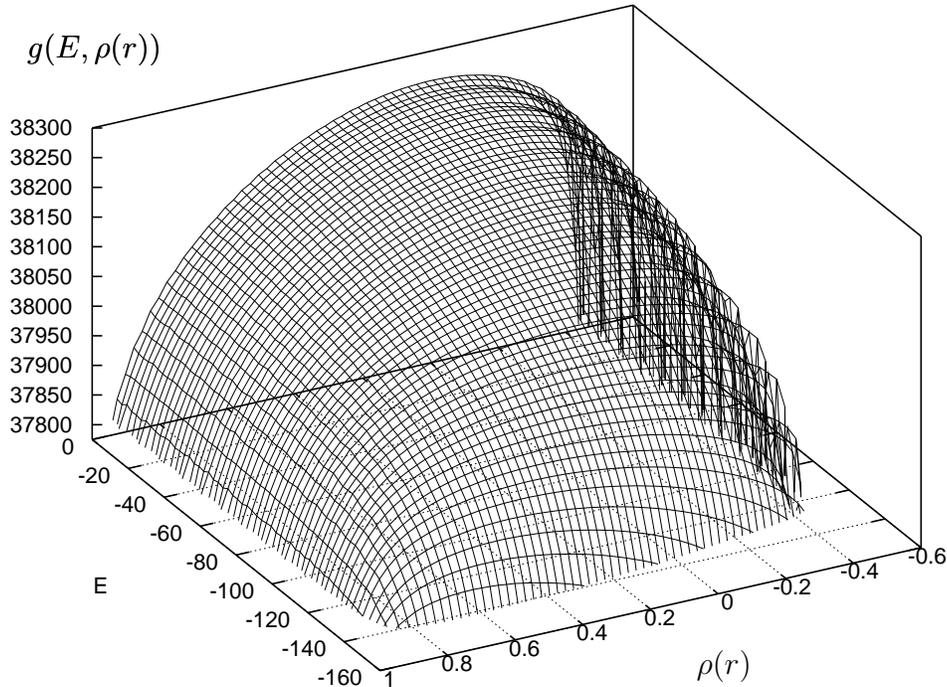}}
\end{center}
\caption{ Three dimensional $g(E, \rho(r))$ surface plotted against energy and
the correlation function for $r$=5.
} 
\label{3d}
\end{figure}

The specific heat per particle has been plotted against temperature Fig. 3 
for $L$=160 and compared with the exact results. In Fig. 4 we have depicted 
the density of state-surface, $\ln \Omega [E,\rho(r)]$ for $r$=5. 
 The scalar order parameter $S$, defined in Eq. (7) has been 
plotted in Fig.5 against temperature for all lattice sizes. It may be 
recalled that the system is disordered at all finite temperature and one would 
expect $S$=0 for all values of $T$.  For a given $T$ (including $T$=0), $S$ rapidly falls 
off with increase in system size, and in the thermodynamic limit will disappear 
altogether, as one would expect due to the finite size effect. 
\begin{figure}[tbh]
\begin{center}
\psfrag{S}{$S$}
\psfrag{T}{$T$}
\psfrag{L}{$L$}
\psfrag{x}{$\big \Downarrow$}
\rotatebox{-90}{\includegraphics[scale=0.6]{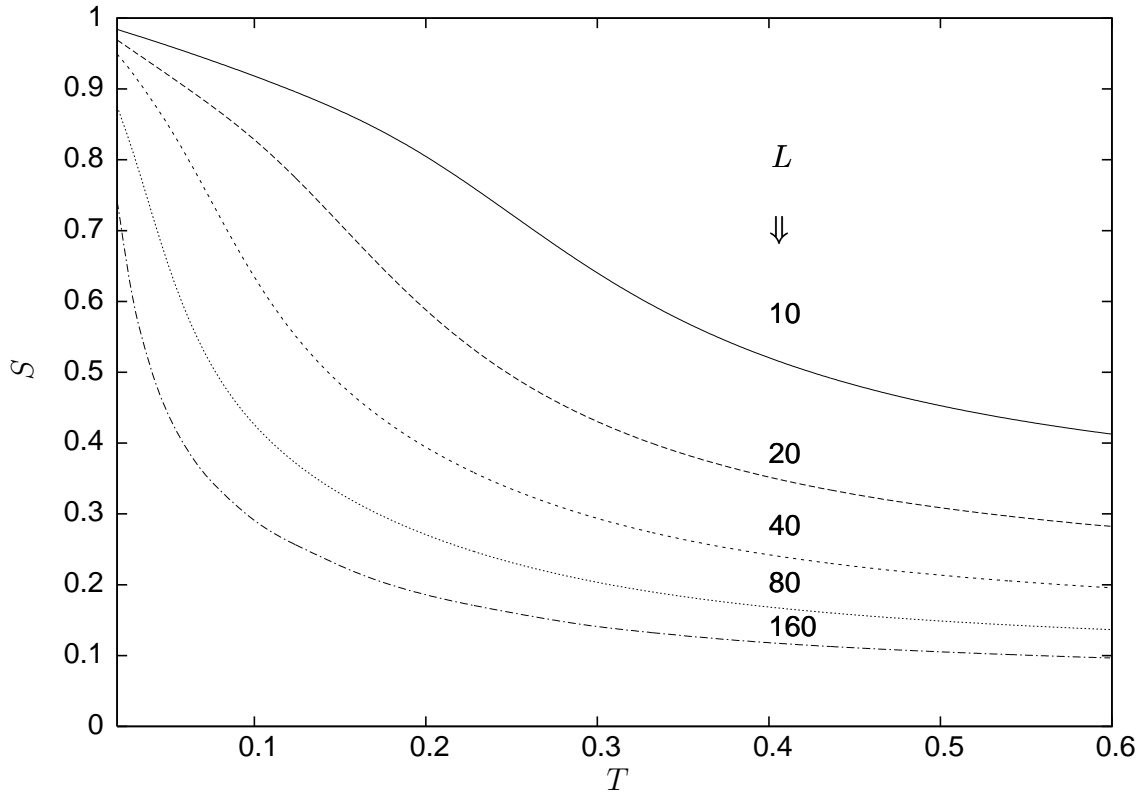}}
\end{center}
\caption{The temperature variation of orientational order parameter for 
different lattice sizes.  
} 
\label{1dorder}
\end{figure}

The correlation function $\rho(r)$ has been compared for the $L$=160 lattice 
using the two-dimensional random walk. We have performed simulations for 12 
values of $r$, ranging from 1 to 40, and these have been plotted against 
temperature in Fig. 6 where comparison has also been made with the exact 
results. It may be noted that, for each value of $r$, we had to run one 
simulation and the joint density of states were determined separately in each 
case. 
The same data has also been shown in Fig. 7 where for three temperatures 
$\rho(r)$ has been plotted against $r$. As one would expect in a disordered 
system, the correlation dies off quickly with increase in $r$. For $r \to 
\infty$, the spins are uncorrelated and $\rho(r)$ should approach $S^2$. 
However, verification of this result from our simulation data will not make 
much sense in a disordered system.  
The CPU time necessary for the $L$=160 lattice for one-dimensional walk is
10.6 hours and the two-dimensional walk involving correlation function is
70.5 hours. For the two-dimensional ($E$,$S$) walk the CPU time is 170 hours.
The program was vectorized ({\it i.e.} the `do' loops parallelized) between two 3.0 GHz Xeon processors in a x226 IBM
Server, automatically by the Intel Fortran Compiler, we used.
\begin{figure}[tbh]
\begin{center}
\psfrag{rho(r)}{$\rho(r)$}
\psfrag{T}{$T$}
\rotatebox{-90}{\includegraphics[scale=0.6]{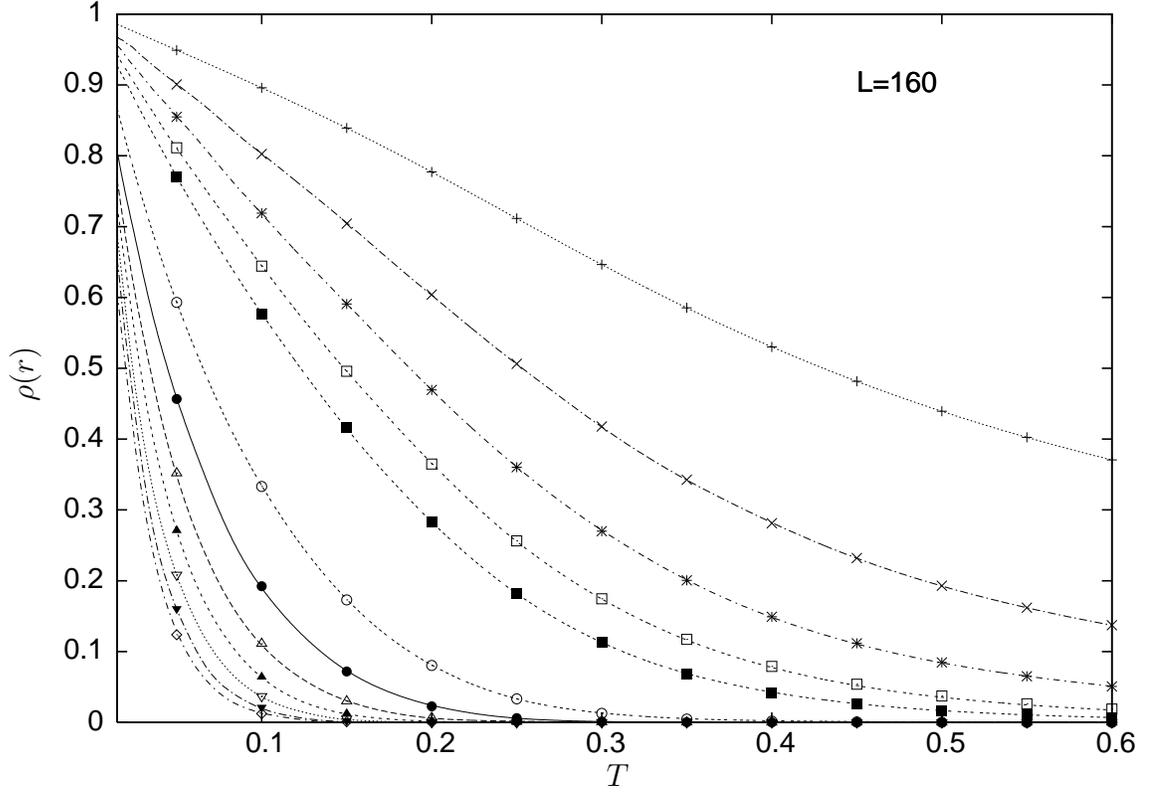}}
\end{center}
\caption{Variation of correlation function $\rho(r)$ with temperature $T$ for
lattice size $L$=160. The points represented by different symbols are from the
exact results obtained in Ref. \cite{exact1,exact2}. The curves are the results
 we obtained from the joint density of states. The values of $r$ taken
are 1, 2, 3, 4, 5, 10, 15, 20, 25, 30, 35 and 40. The topmost curve is for
$r$=1 and the lower curves are for other values of $r$ given in the sequence
above and in ascending order of $r$.
} 
\label{1dcor}
\end{figure}
\begin{figure}[tbh]
\begin{center}
\psfrag{rho(r)}{$\rho(r)$}
\psfrag{r}{$r$}
\psfrag{T}{$T$}
\rotatebox{-90}{\includegraphics[scale=0.6]{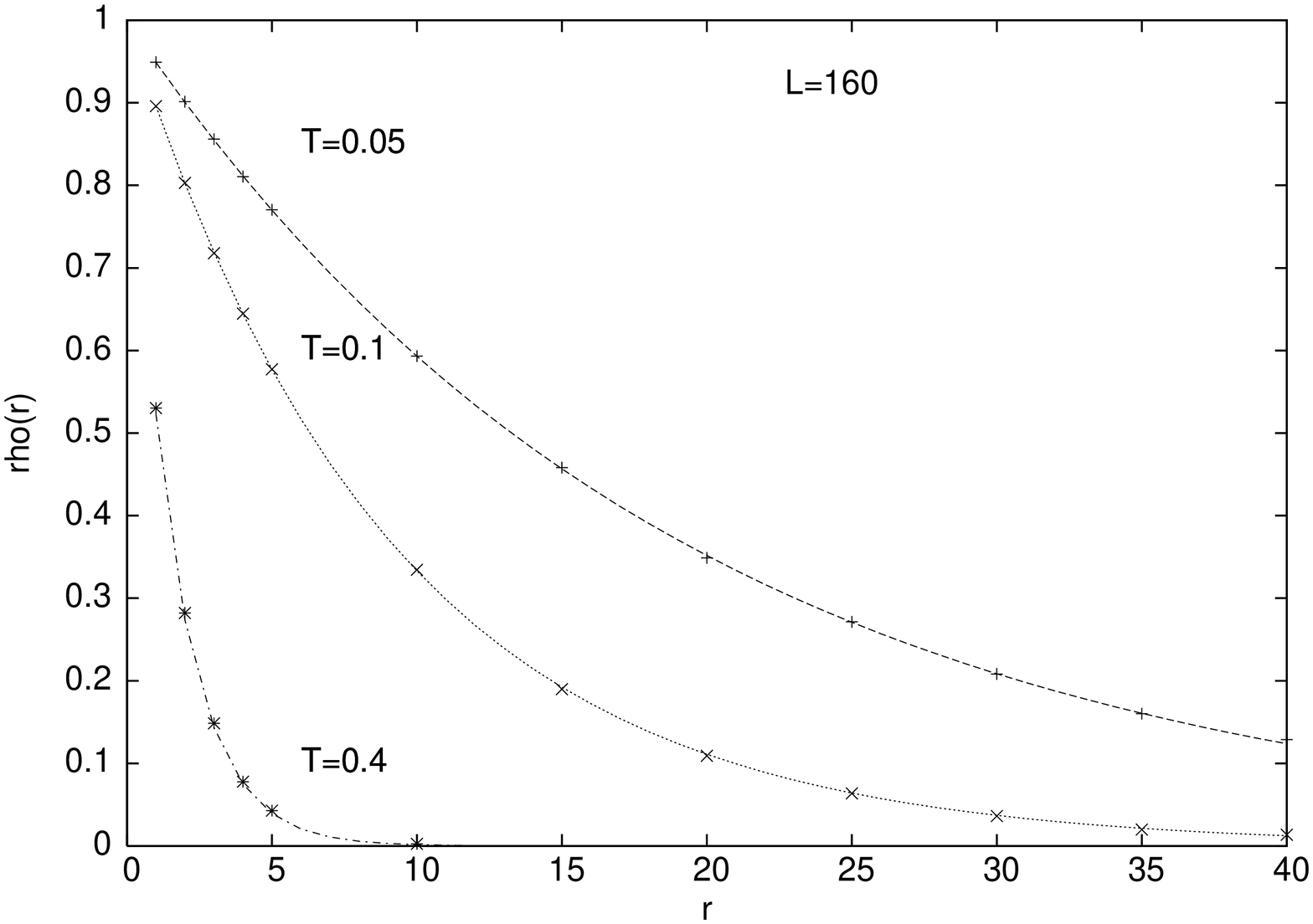}}
\end{center}
\caption{Correlation function as a function of $r$ at different temperatures.}
\label{1dcorr}
\end{figure}

We end this section with a comment on the suggestions of Zhou and Bhatt 
\cite {zhou} and Lee {\it et. al.} \cite{lee}. These  authors have demonstrated 
 that the requirement
 of a flat histogram is not necessary for the convergence of the density of 
states and it is sufficient to have  $1/ \sqrt {\ln f}$ number of samplings of 
each macrostate for  each iteration. They also have tested this 
with WL simulation in a two-dimensional Ising system (ferromagnetic as well as 
the fully frustrated model \cite{lee} ). We have also tried to implement 
this approach, but ended up with rather poor results for the 
two-dimensional random walk. 
 We make a 
general comment that, for a system with a continuous energy spectrum, because 
of 
the infinite number of microstates associated with each bin, the minimum 
number of visits to each bin is expected to be dependent on the bin width and 
this needs a careful consideration for a two-dimensional random 
walk. On the other hand, the flatness criterion 
we have used, works satisfactorily for the determination of joint density 
of states in a continuous model, as our results demonstrate.  
\section{Conclusion}
The work we have presented in this communication, demonstrates that, using a 
flatness criterion one can successfully apply the WL algorithm to simulate a 
continuous model and can get reliable results for the joint density of states. 
We have attempted to provide a guideline for the proper choice of different
parameters involved in the simulation of a continuous model and this may be
useful for further work in such systems.
 However, the two-dimensional random walk in a continuous model becomes
expensive in terms of the CPU time even for a system of moderate size, and 
it seems to be  an impractical task to do this in a reasonably large lattice. 
We have also tried the local and global updating of histograms using Gaussian functions to generate the joint density of states, as has been proposed in the 
work of Zhou {\it et. al.} \cite{2dwalk}, but the results so far have not 
been encouraging. This needs further careful consideration. However, an 
improved or alternative algorithm would perhaps be more helpful.
\clearpage
\noindent{\bf Acknowledgment}
\vskip .1in
We acknowledge the receipt of a research grant No. 03(1071)/06/EMR-II 
from Council of Scientific and Industrial Research, India which helped us to 
procure the IBM x226 servers.

\end{document}